\RequirePackage{fix-cm}
\pdfoutput=1
\documentclass[twocolumn]{svjour3}
\smartqed
\usepackage[table]{xcolor}
\usepackage{graphicx}
\usepackage{epstopdf}
\usepackage{adjustbox}
\usepackage{multirow}
\usepackage{listings}
\usepackage[backend=bibtex, sorting=none]{biblatex}
\usepackage{booktabs,array,dcolumn}
\usepackage{pgfplots}
\usepackage{caption}
\usepackage{wrapfig}
\usepackage{tikz}
\usepackage{amsmath}
\usepackage{subcaption}
\usepackage{adjustbox}
\usepackage{xargs}
\usepackage{color}
\usepackage{soul}
\usepackage{needspace}
\usepackage{amssymb}
\usepackage{amsmath}

\usetikzlibrary{shapes.geometric,positioning}
\usetikzlibrary{lindenmayersystems}
\usetikzlibrary{shapes.multipart}
\usetikzlibrary{automata,arrows}
\usetikzlibrary{petri}
\usetikzlibrary[shadings]
\usetikzlibrary{arrows}
\usetikzlibrary{trees,snakes}
\usetikzlibrary{calc}
\usetikzlibrary{patterns}

\graphicspath{{figures/}}
\pgfplotsset{compat=newest}

\usepackage[colorinlistoftodos,prependcaption,textsize=tiny]{todonotes}
\newcommandx{\change}[2][1=]{\todo[linecolor=red,backgroundcolor=red!25,bordercolor=red,#1]{#2}}
\newcommandx{\suggest}[2][1=]{\todo[linecolor=blue,backgroundcolor=blue!25,bordercolor=blue,#1]{#2}}
\newcommandx{\info}[2][1=]{\todo[linecolor=white,backgroundcolor=teal!25,bordercolor=teal,#1]{#2}}
\newcommandx{\infoblock}[2][1=]{\todo[linecolor=orange,backgroundcolor=orange!25,bordercolor=orange,#1]{#2}}

\bibliography{bib/deep}
\bibliography{bib/parco15}
\bibliography{bib/nmodl}
\bibliography{bib/modeling}
\bibliography{bib/neuron}

\usepackage{etoolbox}
\newcommand*{\affaddr}[1]{#1} 
\newcommand*{\affmark}[1][*]{\textsuperscript{#1}}

\begin{document}
\sloppy

\title{CoreNEURON}
\subtitle{An Optimized Compute Engine for the NEURON Simulator}


\author{Kumbhar, Pramod       \protect\affmark[1]\affmark[*]   \and
        Hines, Michael        \affmark[2]\affmark[*]           \and
        Fouriaux, Jeremy       \affmark[1]                     \and
        Ovcharenko, Aleksandr  \affmark[1]                     \and
        King, James           \affmark[1]                      \and
        Delalondre, Fabien     \affmark[1]                     \and
        Sch\"urmann, Felix     \affmark[1]
}


\institute{
    \textbf{Corresponding author:} \\
    {%
    Felix Sch\"urman,
    Campus Biotech, B1.04,
    Ch. des Mines 9,
    CH-1202 Gen\`eve } \\
    \email{felix.schuermann@epfl.ch} \\
    \affaddr{\affmark[1]Blue Brain Project, \'Ecole Polytechnique F\'ed\'erale de Lausanne (EPFL), Switzerland}\\
    \affaddr{\affmark[2]Yale University, USA}\\
    (*) shared first author \\
}

\date{Received: date / Accepted: date}

\maketitle

\begin{abstract}
The NEURON simulator has been developed over the past three decades and is widely used by neuroscientists to model the
electrical activity of neuronal networks. Large network simulation projects using NEURON have supercomputer allocations
that individually measure in the millions of core hours. Supercomputer centers are transitioning to next generation
architectures and the work accomplished per core hour for these simulations could be improved by an order of magnitude
if NEURON was able to better utilize those new hardware capabilities. In order to adapt NEURON to evolving computer
architectures, the compute engine of the NEURON simulator has been extracted and has been optimized as a library
called CoreNEURON. This paper presents the design, implementation and optimizations of CoreNEURON. We describe how
CoreNEURON can be used as a library with NEURON and then compare performance of different network models on multiple
architectures including IBM BlueGene/Q, Intel Skylake, Intel MIC and NVIDIA GPU. We show how CoreNEURON can simulate
existing NEURON network models with 4-7x less memory usage and 2-7x less execution time while maintaining binary
result compatibility with NEURON.

\end{abstract}

\keywords{
    NEURON, CoreNEURON, Neural Network Simulations, Supercomputing, Performance Optimization
}

\section{Introduction}
 Simulation in modern neuroscientific research has become a third pillar of the scientific method,
 complementing the traditional pillars of experimentation and theory. Studying models of brain components, brain tissue
 or even whole brains provides new ways to integrate anatomical and physiological data and allow insights into causal
 mechanisms crossing scales and linking structure to function. Early studies covered for example the levels from
 channels to cell behavior accounting for detailed morphology (e.g. \cite{schutter:1994}, \cite{Sejnowski:1996}) and
 integrating this detail into models of networks (eg. \cite{traub:1991}). More recently, studies have been accounting
 for increased electrophysiological detail and diversity in the tissue model (e.g. \cite{markram:2015reconstruction},
 \cite{arkhipov:2018}), giving a glimpse at functional importance of the underlying connectome (e.g. \cite{reiman:2017},
 \cite{gal:2017}) allowing for example the reinterpretation of aggregate brain signals such as LFP (e.g.
 \cite{anastassiou:2014}). At the same time, computational studies have strived to look even deeper into the biochemical
 workings of the cell, studying the role of intracellular cascades in neuromodulation (e.g. \cite{lindroos:2018}) or metabolism
 (e.g. \cite{jolivet:2015}), and to abstract some of the detail while maintaining cell type diversity (e.g.
 \cite{Izhikevich3593}, \cite{Potjans2014TheCS}, \cite{dahmen:2016}), or to move the integrated and modeled data all the
 way to fMRI (\cite{deco:2008}).

As the biochemical and biophysical processes of the brain span many orders of magnitudes in space and time, different
simulator engines have been established over time incorporating the appropriate idioms, computational representations
and numerical methods (e.g. at the biochemical level - STEPS~\cite{Schutter:2009:Steps}, at the detailed cellular level
- NEURON~\cite{hines:neuron}, using simplified neuron representations - NEST~\cite{gewaltig2007nest}, or even more
abstract - TVB~\cite{sanzleon:tvb} to name a few).

\begin{table*}
\caption{Summary of Network Models
}
\centering
\begin{tabular}{lp{70mm}lll}

    \toprule
    \multirow{1}{*} {Name}
    & Summary & \#Neurons & \#Compartments  & \#Synapses \\

    \midrule
    \multirow{1}{*} {Traub~\cite{traub:2005}}
    & A single column thalamocortical network model  & 3,560 & 46,5740 & 1,099,820 \\

    \midrule
    \multirow{1}{*} {Dentate~\cite{soltesz:2007}}
    & Dentate Gyrus model including Granule cells with dendritic compartments  & 5,137 & 175,719 & 1,199,988 \\

    \midrule
    \multirow{1}{*} {Ring~\cite{nrnhines-ringtest:github}}
    & Ring network of branching cells & 32,768 & 9,535,488 & 33,280 \\

    \midrule
    \multirow{1}{*} {Cortex + Plasticity~\cite{markram:2015reconstruction}}
    & Somatosensory cortex model with synaptic plasticity  & 219,422 & 99,581,138 &  872,922,040 \\

    \midrule
    \multirow{1}{*} {Hippocampus~\cite{hbp-hippocampus}}
    & Rat Hippocampus CA1 model & 789,595 & 565,495,731 & 361,937,388 \\

    \bottomrule

\end{tabular}
\label{table:network-models}
\end{table*}

The more detail is included in these models and the larger the models become, the larger are the computational
requirements of these simulation engines, making it necessary to embrace advanced computational concepts and faster
computers~\cite{Hepburn2016AccurateRO}~\cite{diesman:2017}~\cite{hines:bgp}.  Table \ref{table:network-models} shows
exemplarily five different network models used in this paper for benchmarking and indicates their size and complexity.
A single-column thalamocortical network model~\cite{traub:2005} is used to better understand population phenomena in
thalamocortical neuronal ensembles. It has 3,560 multi-compartment neurons with soma, branching dendrites and a portion
of axon. It consists of 14 different neuron types, 3,500 gap junctions and 1.1 million connections.  A scaled-down
variant of the full-scale dentate gyrus model~\cite{soltesz:2007} developed in the Soltesz lab~\cite{ivansolteszlab} is
used to understand hippocampal spatial information processing and field potential oscillations.  It consists of 5,143
multi-compartment neurons and 4,121 Poisson spike sources, and includes 6 different cell types, 1.2 million connections
and about 600 gap junctions.  A synthetic model with specific computational characteristics is often needed to evaluate
target hardware based on number of cells, branching patterns, compartments per branch etc. For this purpose, a multiple
ring network model of branching neurons and minimal spike overhead is used~\cite{nrnhines-ringtest:github}. The Blue
Brain Project has published a first-draft digital reconstruction of the microcircuitry of somatosensory cortex in
2015~\cite{markram:2015reconstruction}. This model contains about 219,000 neurons, with 55 layer-specific morphological
and 207 morpho-electrical neuron subtypes. Together with other partners in the European Human Brain Project, this group
is also working on a full-scale model of a rat hippocampus CA1~\cite{hbp-hippocampus}. A first draft of this model
contains about 789,000 neurons with 13 morphological types and 17 morpho-electrical types.

The number of neurons and synapses, however, is not always the best indicator of the computational complexity of a model.
In the  model of \cite{markram:2015reconstruction} each neuron averages to about 20,000 differential
equations to represent its electrophysiology and connectivity. To simulate the microcircuit of 31,000 neurons, it is necessary to
solve over 600 million equations every 25 microseconds of biological time - a requirement far beyond the capabilities of
any standard workstation. It is necessary to utilise massively parallel systems for such simulations but fully
exploiting the capabilities these systems is a challenging task for a large number of scientific codes, including
NEURON. Significant efforts are necessary to prepare scientific applications to fully exploit the massive amount of
parallelism and hardware capabilities offered by these new systems~\cite{exascale-prep:2015}.

In this paper we present our efforts to re-engineer the internal computational engine of the NEURON simulator,
CoreNEURON, to adapt to emerging architectures while maintaining compatibility with existing NEURON models developed by
the neuroscience community. Our work was guided by the goal to leverage the largest available supercomputers for
neuroscientific exploration by scaling the simulator engine to run on millions of threads. A key design goal was to
reduce the memory footprint compared to NEURON as total memory and memory bandwidth are scarce and costly resources when
running at scale. Lastly, for this capability to be easily usable by the normal NEURON community, we endeavored to
tightly integrate CoreNEURON with NEURON.

\section{NEURON Simulation Environment}
\label{section:neuron-simulator}
NEURON is a simulation environment developed over the last 35 years for modeling networks of neurons with complex
branched anatomy and biophysical membrane properties. This includes extracellular potential near membranes, multiple
channel types, inhomogeneous channel distribution and ionic accumulation. It can handle diffusion-reaction models and
integrating diffusion functions into models of synapses and cellular networks. Morphologically detailed models simulated
using NEURON are able to represent the spatial diversity of electrical and biophysical properties of neurons.

Individual neurons are treated as a tree of unbranched cables called \emph{sections}. Each section can have its own set
of biophysical parameters, independently from other sections, and is discretized as a set of adjacent
\emph{compartments} (see e.g.~\cite{hines1993neuron}). Compartmental models of neurons take into account not only the
connectivity between neurons but also the individual morphologies and inhomogeneities of each neuron. The electrical
activity of neurons is modeled using the cable equation (see e.g.~\cite{tuckwell2005introduction}) applied to each
section, where the quantity representing the state of a neuron at a given point in space and instant in time is the
\emph{membrane potential}. The general form of the cable equation for a section, in the case of constant parameters and
conductance based synapse modeling, is given by:

\begin{equation}
\label{eq-cable}
\frac{d}{4R_a }\frac{\partial^2 v}{\partial x^2} = c_m\frac{\partial v}{\partial t} + I_{pas} + I_{ion} + I_{syn}
\end{equation}

where

\begin{list}{$\bullet$}{}
    \item $\displaystyle d \left[\mu m\right] , R_a \left[\Omega cm\right], c_m \left[\frac{\mu F}{cm^2}\right], I_{pas} \left[\frac{mA}{cm^2}\right]$ are biophysical parameters contributing to the passive component of the cable equation (unit conversion factors are not shown but each term has the units of $mA/cm^2$).
    \item $\displaystyle I_{ion} \left[\frac{mA}{cm^2}\right]$ is the active contribution arising from ion channels along the section, whose conductances $g_i$ and resting potentials $e_i$ might depend in a nonlinear fashion upon a set of state variables representing those channels.
    \item $\displaystyle I_{syn} \left[\frac{mA}{cm^2}\right]$ is the contribution from the synapses placed at positions $x_j$, whose conductances $g_j$ and resting potentials $e_j$ might depend in a nonlinear fashion upon a set of state variables and which take effect in a strongly localized manner. Individual synapses have units of $nA$ and conversion to $mA/cm^2$ involves a Dirac delta function, $\delta(x - x_j)$, with units $1/um$, and the diameter; i.e. conversion of absolute current to current per unit area implies division by the compartment area where the synapse is located.
\end{list}

One needs to couple~\eqref{eq-cable} to a set of additional differential equations that describe the evolution of the
states of ion channels and synapses, thus giving rise to a system of PDEs/ODEs as the final problem. Spatial
discretization of the PDEs results in a tree topology set of stiff coupled equations which is most effectively solved by
implicit integration methods. In particular, direct Gaussian elimination with minimum degree ordering is computationally
optimum ~\cite{hines1984efficient}. The general structure of a \emph{hybrid clock-event driven algorithm}
~\cite{hines1993neuron} in NEURON can be divided into a set of operations that are performed at every integration time
step and an interprocess spike exchange operation where a list of spike generation times and identifiers are
synchronized across all processors every minimum spike delay interval. The per integration step operations are :

\begin{list}{$\bullet$}{}
    \item Event-driven spike delivery step where the callback function of each synapse activated by a spike at a given timestep is executed.
    \item Matrix assembly step where the $I_{ion}$ and $I_{syn}$ contributions are computed and included in the matrix.
    \item Matrix resolution step where the membrane potential for the current step is obtained by solving a linear system.
    \item State variables update step where the evolution equations for the states of ion channels and synapses are solved to advance to the current timestep.
    \item Threshold detection step where each neuron is scanned to see if it has met a particular firing condition, and if so a particular list of events is updated.
\end{list}

Although the simulator has demonstrated scaling up to 64,000 cores on the IBM Blue Gene/P system~\cite{hines:bgp}, with
the emerging computing architectures (like GPUs, many-core architectures) the key challenges are numerical efficiency
and scalability. The simulator needs to : 1) expose fine grain parallelism to utilize the massive number of hardware
cores, 2) be optimized for memory hierarchies and 3) fully utilize processor capabilities such as vector units. To
simulate models with billions of neurons on a given computing resource, memory capacity is another major challenge. In
order to address these challenges, the compute algorithm of the NEURON simulator was extracted and optimized into a
standalone library called CoreNEURON.


\section{CoreNEURON Design and Implementation}
The integration interval operations (listed in Section \ref{section:neuron-simulator}) consume most of the simulation
time~\cite{Kumbhar:deep-isc2016}. The goal of CoreNEURON is to efficiently implement these operations considering
different hardware architectures. This section describes the integration of CoreNEURON with the NEURON execution
workflows, major data structure changes to reduce memory footprint, memory transfer between NEURON-CoreNEURON and
a checkpoint-restore implementation to facilitate long running simulations.

\subsection{NEURON to CoreNEURON Workflow}

\begin{figure*}[t]
  \centering
  \includegraphics[width=1\textwidth,keepaspectratio]{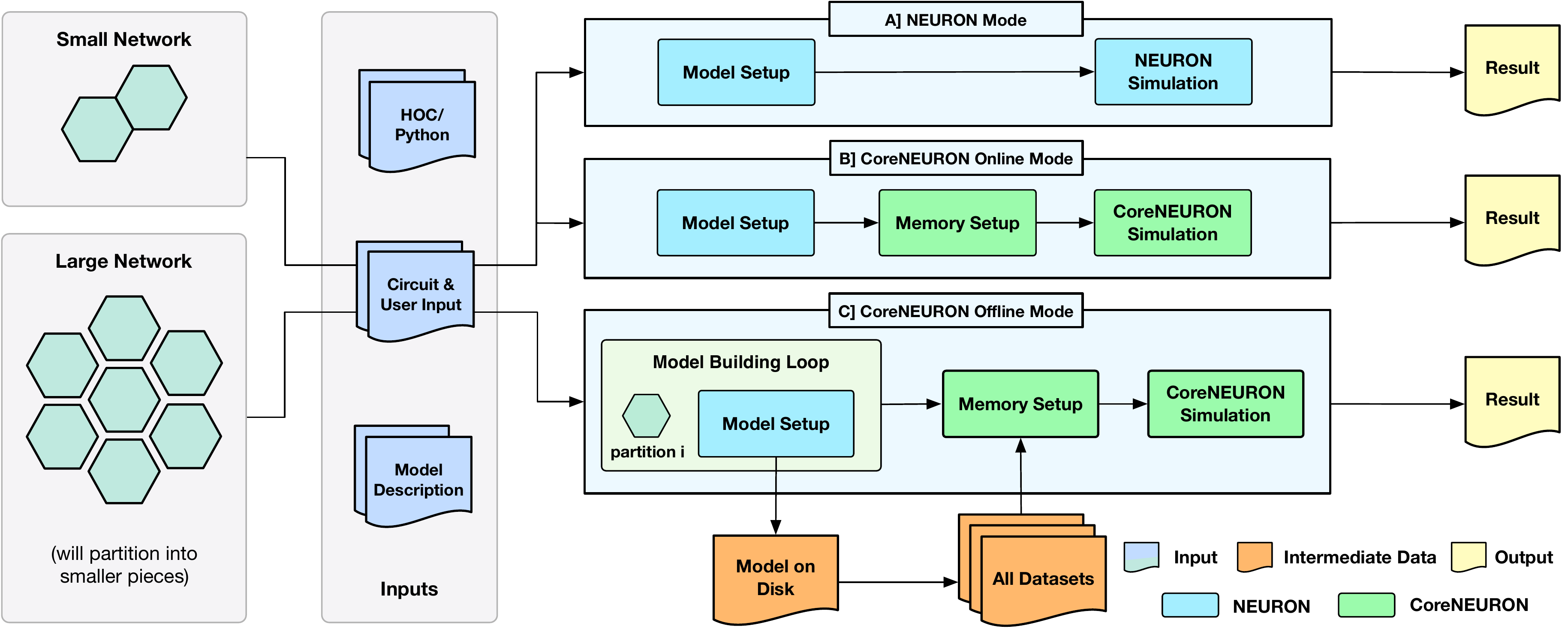}
  \caption{ Different execution workflows supported by NEURON simulator with CoreNEURON : A] shows existing simulation workflow used by all NEURON users; B] shows new CoreNEURON based workflow where in-memory model is transferred using direct memory access and then simulated by CoreNEURON; C] shows new CoreNEURON based workflow where NEURON partition large network model into smaller chunks, build in-memory model iteratively and then CoreNEURON loads whole model for simulation}
  \label{fig:neuron-coreneuron-workflow}
\end{figure*}

One of the key design goal of CoreNEURON is to be compatible with the existing NEURON models and user workflows. With
the integration of CoreNEURON library, the NEURON simulator supports three different workflows depicted in
Figure~\ref{fig:neuron-coreneuron-workflow}.

\begin{list}{$\bullet$}{}
\item NEURON mode
\item CoreNEURON Online mode
\item CoreNEURON Offline mode
\end{list}

Existing users are familiar with the default \emph{NEURON mode}. The model descriptions written in
NMODL~\cite{nmodl:hines2000} are used to build a dynamically loadable shared library. The HOC/Python scripting interface
is used to build network model in memory (\emph{Model Setup phase}). This in-memory model is then simulated using the
\emph{hybrid clock-event driven algorithm} described in Section \ref{section:neuron-simulator} (\emph{Simulation
phase}). Users have full control over model structure and can introspect or record all events, states, and model
parameters using the scripting or graphical user interface (\emph{Result phase}).

\emph{CoreNEURON Online Mode} allows users to run their models efficiently with minimal changes. After the \emph{Model
Setup phase}, the in-memory representation is copied into CoreNEURON's memory space. CoreNEURON then re-organizes the
memory during \emph{Memory Setup phase} for efficient execution (see Section
\ref{subsection:data-layout-vectorisation}).  The \emph{Simulation phase} is executed in CoreNEURON and spike results
are written to disk. Note that the same NMODL model descriptions are used both in NEURON as well as CoreNEURON.

\emph{CoreNEURON Offline} mode is intended for large network models that cannot be simulated with NEURON due to memory
capacity constraints. In this mode, instead loading the entire model at once, the \emph{Model Setup phase} builds a subset
of the model that fits into available memory. That subset is written to disk, the memory used by the subset is freed,
and the \emph{Model Setup phase} constructs another subset. After all subsets are written by NEURON, CoreNEURON reads
the entire model from the disk and begins the \emph{Simulation phase}. Because CoreNEURON's cell and network connection
representations are much lighter weight than NEURON's, 4-7x larger models than NEURON can be simulated with CoreNEURON
(see Section \ref{section:models-performance}).

Users can adapt existing models to the \emph{CoreNEURON Online Mode} workflow with the trivial replacement of the
\emph{psolve} function call with \emph{nrncore\_run} of the ParallelContext class~\cite{nrn:parallelcontext}. Presently,
however, event callbacks from CoreNEURON to NEURON interpreter code are not implemented (see Section
\ref{section:conclusion-future-work}).

\begin{table*}
\caption{Memory footprint comparison for different data structures (in bytes)}
\centering
\begin{tabular}{lp{70mm}ll}

    \toprule
    \multirow{1}{*} {Data Structure}
    & Purpose & NEURON & CoreNEURON \\

    \toprule

    \multirow{1}{*} {Node}
    & Compartment of the neuron   & 128 & - \\

    \midrule
    \multirow{1}{*} {Section}
    & Unbranched cable of the neuron  & 96 & - \\

    \midrule
    \multirow{1}{*} {Object}
    & High level HOC object  & 64 & - \\

    \midrule
    \multirow{1}{*} {Presyn}
    & Synapse object at origin & 208 & 64\\

    \midrule
    \multirow{1}{*} {InputPresyn}
    & Similar to Presyn  & - & 24 \\

    \midrule
    \multirow{1}{*} {Point\_process}
    & Synapse overhead  & 56 & 8 \\

    \midrule
    \multirow{1}{*} {Prop}
    & Property object in compartment  & 48 & - \\

    \midrule
    \multirow{1}{*} {Netcon}
    & Connection between neuron & 56 & 40 \\

    \midrule
    \multirow{1}{*} {Pointer}
    & Memory address & 8 & 4 \\

    \midrule
    \multirow{1}{*} {Memb\_list}
    & List of mechanisms or channels  & 56 & 64 \\

    \midrule
    \multirow{1}{*} {NrnThreadMembList}
    & Mechanism list for group of neurons  & 34 & 40 \\

    \midrule
    \multirow{1}{*} {PreSynHelper}
    & Helper object for PreSyn  & - & 4 \\

    \midrule
    \multirow{1}{*} {Symbol}
    & Token parsed by HOC interpreter & 56 & - \\


    \bottomrule
\end{tabular}
\label{table:data-structure-memory-sizes}
\end{table*}

\subsection{Data Structure Changes}
\label{subsection:data-structure-changes}

NEURON is used as a general framework for designing and experimenting with neural models of varying anatomical detail
and membrane complexity. Users can interactively create cells with branches of varying diameters and lengths, insert
ionic channels, create synapses, record and visualize different properties using a GUI. In order to provide this
flexibility and introspection capability for examining a large number of complex data structures are created. But, once
the users are satisfied with the behavior of the model, they run larger/longer simulations on workstations or clusters
where interactivity or detailed introspection capabilities are often not required. In this type of batch execution,
memory overhead from many large, complex data structures with many mutual pointers can be significantly reduced by
replacing them with fixed arrays of data structures in which the few necessary pointers are replaced by integers. For
example, the network connection object (\emph{Netcon}) and the common synapse base class (\emph{Point\_process}), which
are responsible for a significant portion of memory usage in NEURON, were reduced from 56 to 40 bytes and 56 to 8 bytes
respectively in CoreNEURON. Table \ref{table:data-structure-memory-sizes} lists the important data structures and their
memory usage comparison between NEURON and CoreNEURON. CoreNEURON eliminates the Python/HOC interpreter and so, data
structures like \emph{Node}, \emph{Section}, \emph{Object} are no longer needed. The memory usage
improvements from these optimizations for different network models are discussed in Section \ref{section:models-performance}.


\subsection{Pointer Semantics}

NEURON users can define their own data structures and allocate memory through the use of \emph{POINTER} and
\emph{VERBATIM} constructs of NMODL~\cite{nrn:nmodluserguide}. Many internal data structures of NEURON use pointer
variables to manage various dynamic properties, connections, event queues etc. As a model is built incrementally using
the scripting interface, various memory pools are allocated during the \emph{Model Setup phase}. As data structures
between NEURON and CoreNEURON are different, serializing memory pools
becomes one of the major memory management challenges of the CoreNEURON implementation. With serialization, pointer
variables need to be augmented with meta information to allow proper decoding by CoreNEURON. This meta information
indicates the \emph{pointer semantics}.  All data variables which potentially are the target pointers are grouped into a
contigiuous memory pool and pointer variables are converted to an integer offset into the memory pool. When the NEURON
pointers are copied to CoreNEURON's memory space, the semantic type associated with the pointer variable is used to
compute the corresponding integer offset. Table \ref{table:semantics} enumerates the different semantics types
introduced to facilitate memory serialization. For example, \emph{area} represents compartment area (\emph{8 byte}
double) but \emph{pntproc} represents the larger \emph{Point\_process} object and hence needs a different decoding
mechanism.

\begin{figure*}[t]
    \centering
    \includegraphics[width=1\textwidth]{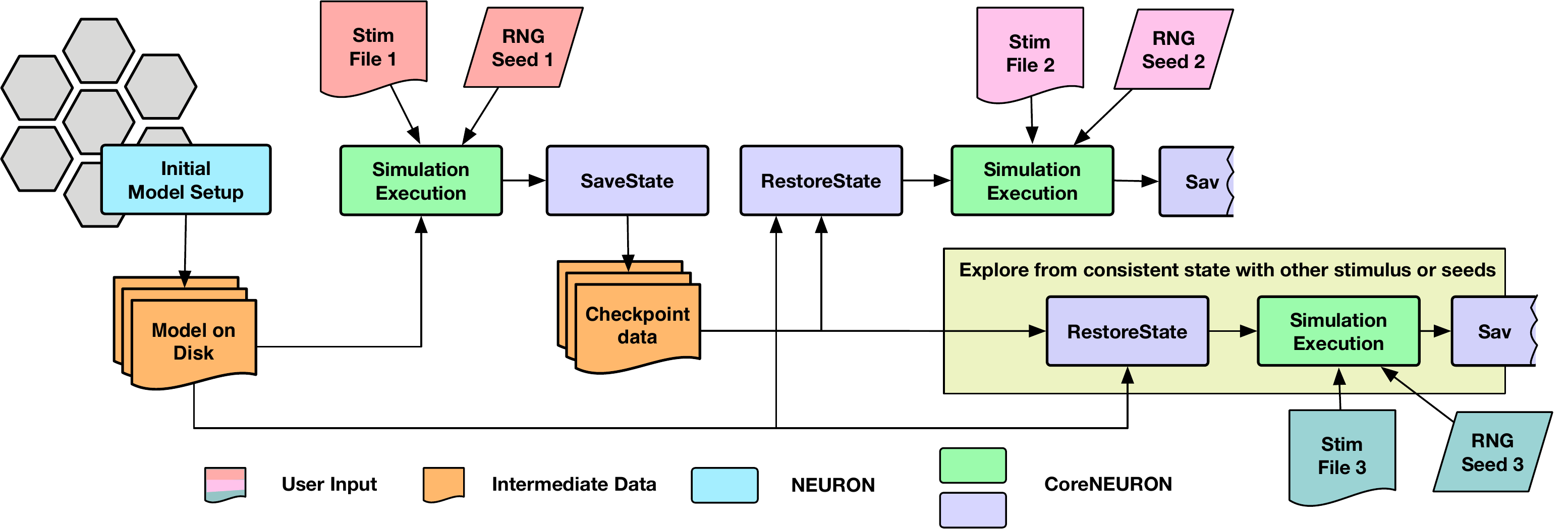}
    \caption{Simulation workflow with the checkpoint-restart feature with the flexibility to re-launch simulations with different stimuli or random
number streams}
    \label{fig:savestate-resume}
\end{figure*}

\subsection{Checkpoint-Restart Support}

The network simulations for studying synaptic plasticity can run from a week to a month. Enabling such simulations of long
biological time-scales is one of the important use cases for CoreNEURON. Most of the cluster and supercomputing resources
have a maximum wall clock time limit for a single job (e.g. up to 24 hours). The checkpoint-restart~\cite{schulz:2004} is
commonly used technique to enable long running simulations and has been implemented in CoreNEURON. Since the checkpoint 
operation could take place at anytime with varying degrees of cell firing activity, it was necessary to account for generated yet undelivered
 synaptic events in addition to saving the in-memory state of the simulator. When a cell fires, it may
have many connections to other cells with different delivery delays. During the checkpoint operation, any undelivered messages
are collapsed back into the original event of the firing cell so that a single event can be saved. Once the network
simulation is checkpointed, users have flexibility to launch multiple simulations with different stimuli or random
number streams in order to explore network stability and robustness. The execution workflow of such simulations is shown
in Figure~\ref{fig:savestate-resume}.

\begin{table}
\caption{Semantic type and their purpose}
\centering
\begin{tabular}{lll}
    \toprule
    \multirow{1}{*} {Semantic name}
    & Purpose \\

    \toprule
    \multirow{1}{*} {area}
    & area of the compartment \\

    \midrule
    \multirow{1}{*} {iontype}
    & type of ion used (ca, na, k)  \\

    \midrule
    \multirow{1}{*} {cvodeieq}
    & element on event queue  \\

    \midrule
    \multirow{1}{*} {netsend}
    & network send event \\

    \midrule
    \multirow{1}{*} {pointer}
    & pointer used in NEURON  \\

    \midrule
    \multirow{1}{*} {pntproc}
    & point process  \\

    \midrule
    \multirow{1}{*} {corepointer}
    & pointer used in CoreNEURON  \\

    \midrule
    \multirow{1}{*} {watch}
    & element used as watch statement  \\

    \midrule
    \multirow{1}{*} {diam}
    & diameter of the compartment \\
    \bottomrule
\end{tabular}
\label{table:semantics}
\end{table}

\subsection{Portability Considerations}

CoreNEURON can transparently handle all spiking network simulations including gap junction coupling with the fixed time
step method. The model descriptions written in NMODL need to be thread safe~\cite{nrn:parallelcontext-thread} to exploit
vector units of modern CPUs and GPUs. This can be achieved with the help of NEURON's \emph{mkthreadsafe} tool. New keywords like
\emph{COREPOINTER} and \emph{CONDUCTANCE} have been added to NMODL to facilitate serialization and improve performance
optimization respectively. These keywords are also backported to NEURON so that the models remain compatible for either
NEURON or CoreNEURON execution.  For scalability and portability of random numbers on platforms like GPUs, CoreNEURON
supports the Random123 pseudo-random generator~\cite{Salmon:2011ran123}.

\section{Optimizations}
\label{section:optimisation}
In order to improve the performance of CoreNEURON on different architectures, different optimization schemes are
implemented for multi-threading, memory layout, vectorization and code generation. These optimizations are described in
this section.

\begin{figure*}[t]
  \centering
  \includegraphics[width=0.9\textwidth,keepaspectratio]{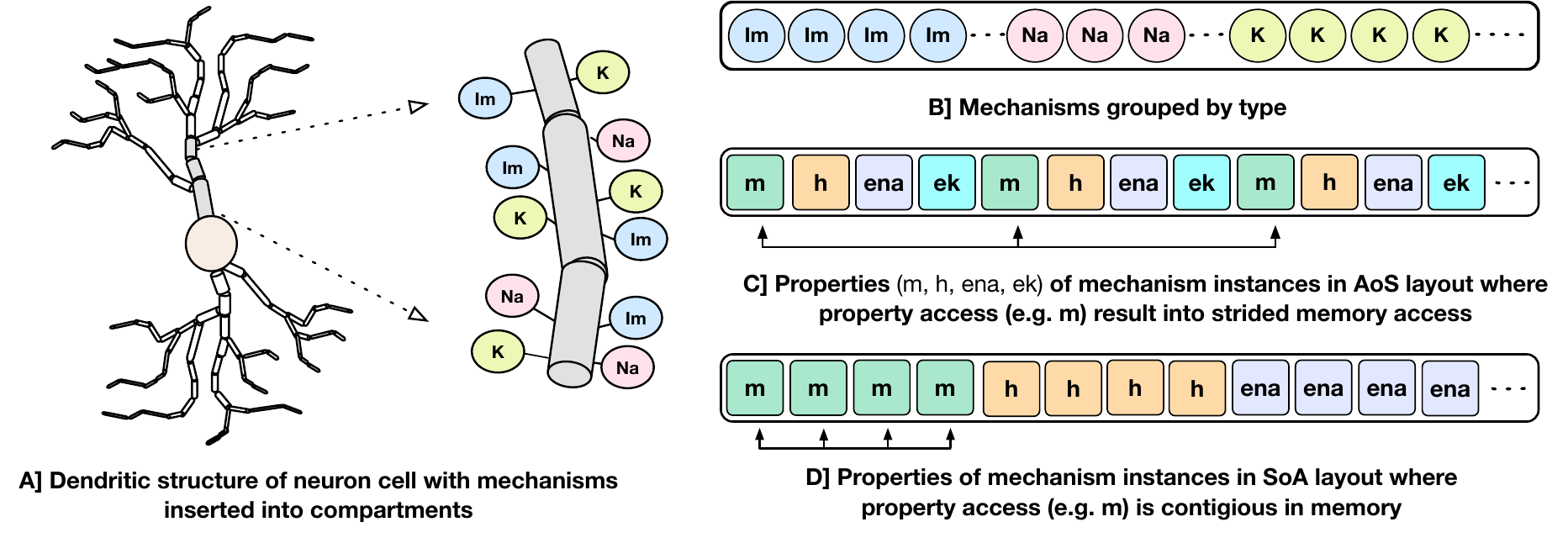}
  \caption{ A schematic representation of dendritic structure of a neuron with different mechanisms inserted into the compartment is shown on the left. On the right: B] shows how NEURON and CoreNEURON groups the
  mechanism instances of the same type; C] shows how NEURON stores properties of individual mechanism in the \emph{AoS} layout; D] shows the new \emph{SoA} layout in CoreNEURON for storing mechanism properties}
  \label{fig:cell_channels}
\end{figure*}

\subsection{Parallelism} \label{subsection:thread-parallelsim}

Both NEURON and CoreNEURON use the \emph{Message Passing Interface} (MPI) to implement distributed memory parallelism.
Although NEURON supports multi-threading based on Pthread~\cite{multithreading:1996}, users
commonly use pure MPI execution due to better scaling behavior. But, pure MPI execution will affect scalability due to
MPI communication and memory overhead of internal MPI buffers when executing at scale~\cite{Lange2013AchievingES}.
To address this scalability and parallelism challenge, CoreNEURON relies on three distinct level of parallelism. First,
at the highest level, a set of neurons that have equivalent computational cost are grouped together and assigned to each
MPI rank on the compute node. Second, within a node, an individual neuron group is assigned to an OpenMP~\cite{openmp:1998}
thread executing on a core. This thread simulates the given neuron group for the entire simulation ensuring data
locality. Finally, vector units of the core are utilized for executing groups of channels in parallel. This allows
simulations with a single MPI process per compute node to meet the scalability challenge.

\subsection{Memory Layout and Vectorization} \label{subsection:data-layout-vectorisation}

Processor memory bandwidth is one of the scarce resources and often the major impediment to improve the performance of
many applications including NEURON. The compute kernels of channels and synapses are bandwidth limited and can reach
close-to-peak memory bandwidth~\cite{deep:Kumbhar2016}. The dendritic structures of a neuron are divided into small
compartments and different membrane channels or mechanisms are inserted into different compartments
(Figure~\ref{fig:cell_channels}A). For memory locality, both NEURON and CoreNEURON groups the channels by their type as
shown in Figure~\ref{fig:cell_channels}B. But, NEURON organizes properties of individual mechanisms (like \emph{m},
\emph{h}, \emph{ena}) in the \emph{Array of Structs (AoS)} memory layout (Figure~\ref{fig:cell_channels}C).  When a
specific property is accessed, for example, \emph{m}, it results in strided memory accesses with inefficient memory
bandwidth utilization and hence poor performance. To address this issue, CoreNEURON organizes channel properties into
the \emph{Structure of Arrays (SoA)} memory layout (Figure~\ref{fig:cell_channels}D). This allows efficient compiler
vectorization and efficient memory bandwidth utilization for all channel and synapse computations. The performance
improvements from this optimization is discussed in~\cite{deep:Kumbhar2016}.

\subsection{NMODL Source-to-Source Translator}
\label{subsection:mod2c}

\begin{figure*}
  \centering
  \includegraphics[width=1\linewidth]{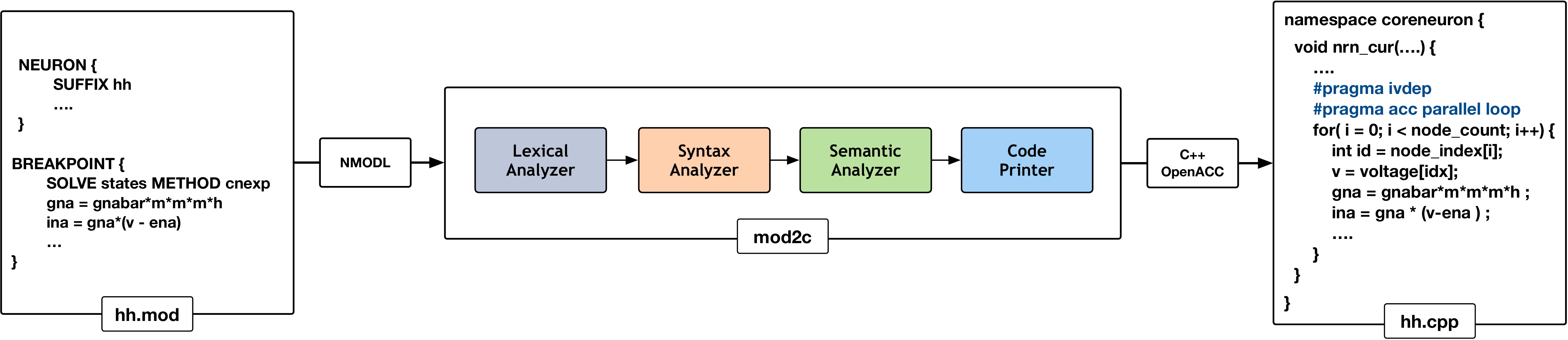}
  \caption{Code generation workflow for CoreNEURON : different phases of source-to-source compiler are shown in the middle that translates model description file (hh.mod)  to C++ code (hh.cpp) and inserts compiler hints for CPU/GPU parallelization}
  \label{fig:nmodl_lexer_parser}
\end{figure*}

NEURON has had support for code generation through the model description language, NMODL, since version 2 released in
1989~\cite{codegen-inga:2018}. The code generation program of NEURON has been modified into a standalone tool called
\emph{MOD2C}~\cite{mod2c:github}. This tool is used by CoreNEURON to support all NEURON models written in NMODL.
Figure~\ref{fig:nmodl_lexer_parser} shows the high level workflow of MOD2C. The first step of source-to-source
translator is \emph{lexical analysis} where lexical patterns in the NMODL code are detected and tokens are generated.
The \emph{syntax analysis} step uses those tokens and determine if the series of tokens are appropriate in the language.
The \emph{semantic analysis} step make sure if syntactically valid sentences are meaningful as part of the model
description. \emph{Code generation} is the step in which a C++ file is created with compiler hints for auto-vectorization
and GPU parallelization with the OpenACC programming model ~\cite{OpenACC:openaccorg}. MOD2C also takes care of code
generation for \emph{AoS} and \emph{SoA} memory layouts.  MOD2C uses open source flex and bison tools~\cite{Levine:2009}
for this implementation. More information about the NMODL code generation pipeline can be found in
~\cite{codegen-inga:2018}.

\subsection{GPU Porting}

Prior to the CoreNEURON project, a substantial effort was made to port NEURON to the GPU architecture using the CUDA
programming model~\cite{Nickolls:2008}. One of the two major components of this implementation was the extension of the
NMODL source-to-source compiler to emit CUDA code. The other major component managed an internal memory transformation
from NEURON's thread efficient \emph{AoS} memory layout to a more GPU memory efficient \emph{SoA} layout. This
experimental NEURON version~\cite{nrn-gpu:bitbucket} was quite efficient for matrix setup and channel state integration
for cellular simulations but did not reach network simulation capability. The project foundered on software
administration difficulties of maintaining two completely separate codebases, the difficulty of understanding the data
structure changes involved for memory layout transformation from \emph{AoS} to \emph{SoA}, and the difficulty of
managing pointer updates in the absence of pointer semantics information. It became clear that a more general view was
required that could not only alleviate these problems for the GPU but had a chance of evolving to work on future
architectures. This view is embodied in CoreNEURON development. As discussed in Section
~\ref{subsection:data-layout-vectorisation}, CoreNEURON data structures and memory layout have been optimized for
efficient memory access. MOD2C supports code generation with the OpenACC programming model that helps to target
different accelerator platforms. Users need to compile the CoreNEURON library with a compiler that supports OpenACC
(e.g. PGI, Cray).

\begin{figure*}[t]
  \centering
  \includegraphics[width=0.95\textwidth,keepaspectratio]{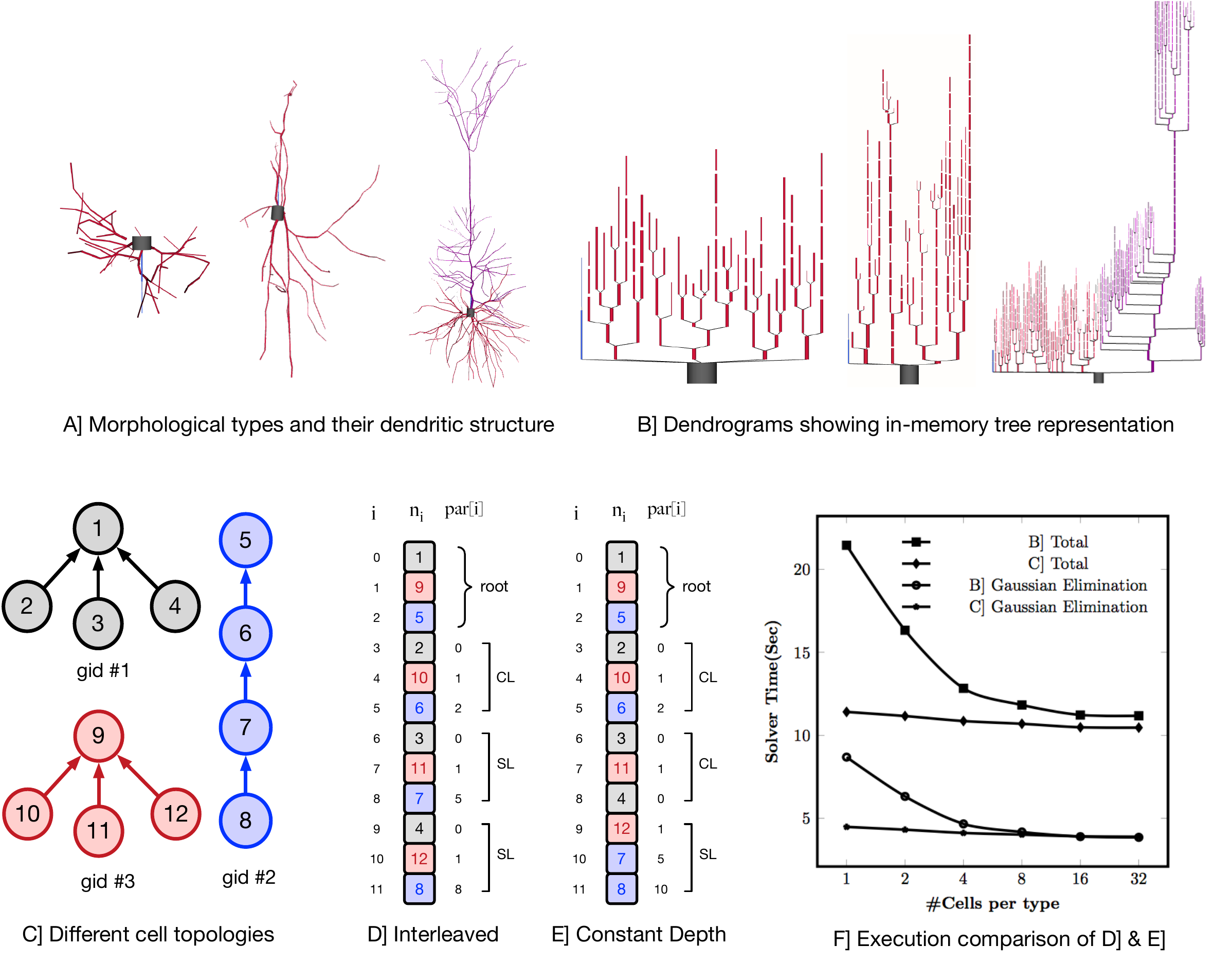}
  \caption{The top row shows three different morphological types with their dendritic tree structure in A] and dendrograms showing in-memory tree representation of these types in CoreNEURON in B]. The bottom row shows different node ordering schemes to improve the memory access locality on GPUs : C] Example topologies of three cells with the same number of compartments; D] Interleaved Layout where a compartment from each of N cells forms an adjacent group of N compartments. For \emph{i}th node, \emph{n$_{\text{i}}$} is node index and \emph{par[i]} is its parent index. With three executor threads, square brace highlight parent indices that result into contiguous memory load (CL) and strided memory load (SL); E] Constant Depth Layout where all nodes at same depth from root are adjacent; F] Comparison of two node ordering schemes for Ring network model showing execution time of whole simulation and Gaussian Elimination step.}
  \label{fig:morphological-diversity-ordering}
\end{figure*}

One of the performance challenges for a GPU implementation is irregular memory accesses due to the non-homogeneous tree
structure of neurons. For example, Figure~\ref{fig:morphological-diversity-ordering}A shows three different
morphological types and their compartmental tree connection topology in the simulator is shown in
Figure~\ref{fig:morphological-diversity-ordering}B. The GPU delivers better performance when consecutive threads (in
groups of 16 or 32) perform the same computations and load the data from consecutive memory addresses. When there are a
large number of cells per morphological type, it is straightforward to achieve optimal performance by interleaving the
compartments of identical cells. But, with few cells per morphological type, Gaussian elimination suffers from
non-contiguous layout of parents relative to a group of nodes. This results in irregular, strided memory accesses and
hence poor performance~\cite{cuHinesBatch:2017}. To address this, two alternative node orderings schemes,
\emph{Interleaved} layout and \emph{Constant Depth} layout, are illustrated in
Figure~\ref{fig:morphological-diversity-ordering}D and Figure~\ref{fig:morphological-diversity-ordering}E. All cells
have the same number of compartments but each has a different branching pattern. Nodes (representing compartments) within
a cell are numbered with successive integers. In the case of \emph{Interleaved} layout, a compartment from each of \emph{N}
cells forms an adjacent group of \emph{N} compartments. The groups are in any root to leaf order but corresponding
compartments in identical cells are adjacent. As an example, for a group of three threads the vertical square braces
highlight parent indices that have the same order as the nodes. This results in either contiguous memory loads
(\emph{CL}) or strided memory load (\emph{SL}). For each Gaussian elimination operation the number of threads that can
compute in parallel is equal to the number of cells and hence this scheme is referred as \emph{one cell per thread
layout}. For \emph{Constant Depth} layout, all nodes at the same depth from the root are adjacent. For a given depth,
corresponding nodes of identical cells are adjacent. Children of branch nodes in the same cell are kept as far apart as
possible to minimize contention while updating the same node from different threads.

To analyse the impact of node ordering schemes on the execution time, we used a multiple Ring network model of cells with random tree
topology~\cite{nrnhines-ringtest:github}. This test allows to evaluate performance impact when parents of a
contiguous group of 32 nodes are not contiguous and executed by a 32 thread (\emph{warp}). We used a multiple Ring model
with a total of 131,072 cells comprising 10,878,976 nodes running for 10ms on NVIDIA K20X GPU~\cite{tesla-k20x}. Every
cell has the same number (83) of nodes but different cell types have a different random branching pattern of the 40
dendrites. The number of identical cells per type ranges from 1 (131,072 distinct branching patterns) to 32 (4096
distinct branching patterns). Note that regardless of the branching pattern, Gaussian elimination takes exactly the same
number of arithmetic operations.  Figure~\ref{fig:morphological-diversity-ordering}D shows performance of Interleaved Layout and
Constant Depth Layout. For both node ordering schemes, performance is optimal with regard to parent ordering when there
are at least 32 cells of each type corresponding to the 32 threads operating in \emph{Single Instruction Multiple Data
(SIMD)} mode. With fewer cells per type, parent node ordering becomes less than optimal and the performance of
Interleaved layout suffers by up to a factor of two. Note that the total runtime deteriorates more rapidly than Gaussian
elimination time due to the fact that the parent contiguity also affects the performance of tree matrix setup during
evaluation of a node's current balance equation. The execution time of \emph{Constant Depth} layout shows that it is
possible to permute node ordering so that parent nodes are more likely to be in significant contiguous order relative to
their children. The constant ratio between total runtime and Gaussian elimination is due to negligible time contribution
of passive dendrites to matrix setup in combination with the significant role of parent ordering in computing the effect
of topologically adjacent nodes on matrix setup of the current balance equations.

\begin{table*}
\caption{Details of Benchmarking Systems}
\centering
\begin{tabular}{lll}

    \toprule
    \multirow{3}{*} {BlueGene/Q (BB4)}
    & Processor              & IBM PowerPC A2, 16 cores @ 1.6 GHz, 16 GB DRAM \\
    & Compiler toolchain     & IBM XL 12.1 and IBM MPI \\
    & Network                & Integrated 5-D torus \\

    \midrule
    \multirow{3}{*} {Intel Skylake (BB5)}
    & Processor              & 2 Xeon 6140, 36 cores @ 2.3 GHz, 384 GB DRAM \\
    & Compiler toolchain     & Intel 2018.1 and HPE-MPI (MPT) \\
    & Network                & InfiniBand EDR \\

    \midrule
    \multirow{4}{*} {Intel KNL (BB5)}
    & Processor              & Xeon Phi (7230), 64 cores @ 1.3 GHz, 16 GB MCDRAM, 96 GB DRAM  \\
    & Compiler toolchain     & Intel 2018.1 and HPE-MPI (MPT) \\
    & Network,               & InfiniBand EDR \\

    \midrule
    \multirow{4}{*} {NVIDIA GPU (BB5)}
    & Processor              & NVIDIA GPU V100 SXM2  \\
    & Compiler toolchain     & PGI 18.10, OpenMPI 2.0\\
    & Network                & InfiniBand EDR \\

    \bottomrule

\end{tabular}
\label{table:benchmark-systems}
\end{table*}

\section{Benchmarks and Performance}
\label{section:models-performance}

Not all network models are compute intensive or benefit equally from CoreNEURON optimizations. In order to evaluate the
performance improvements with the optimizations discussed in the previous section we ran several published network
models listed in Table\ref{table:network-models} on different computing architectures. This section describes the
benchmarking platforms and compares performance between NEURON and CoreNEURON.

The benchmarking systems with hardware details, compiler toolchains and network fabrics are summarized in Table
 \ref{table:benchmark-systems}. The Blue Brain IV (BB4) and Blue Brain V (BB5) systems are based on IBM
BlueGene/Q~\cite{bgq:2012} and HPE SGI 8600~\cite{hpe-sgi-8600} platforms respectively, hosted at the Swiss National
Computing Center (CSCS) in Lugano, Switzerland. The BB4 system has 4,096 nodes comprising 65,536 PowerPC A2 cores. The
BB5 system has three different compute nodes: Intel KNLs with low clock rate but high bandwidth MCDRAM, Intel Skylakes
with high clock rate, and NVIDIA Volta GPUs. Vendor provided compilers and MPI libraries are used on both systems. The
BB4 system is used for strong scaling benchmarks as it has a large core count compared to BB5 system.

\begin {figure*}[t]
    \centering
    \includegraphics[width=1\textwidth]{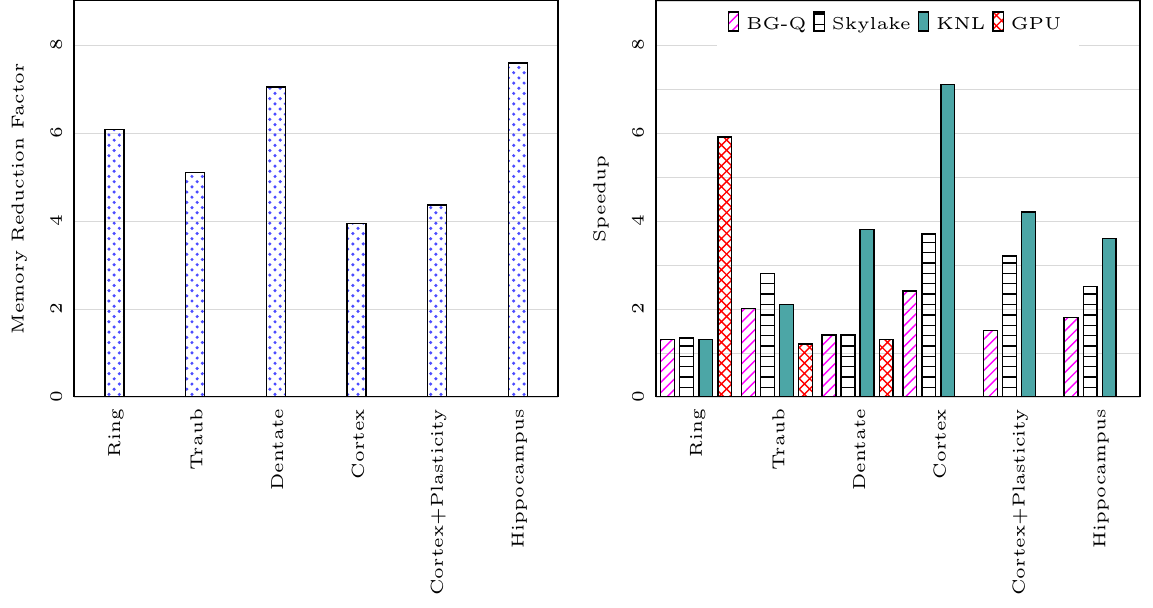}
    \caption {Memory usage reduction with CoreNEURON compared to NEURON (measured on BB4) on the left and speedup of CoreNEURON simulations compared to NEURON on various architectures on the right}
    \label{fig:memory-reduction}
\end{figure*}

We compared the memory footprint of different network models listed in Table \ref{table:network-models}.
Figure~\ref{fig:memory-reduction} on the left shows memory usage reduction with CoreNEURON simulation compared to NEURON
simulation. The memory reduction factor depends on various model properties (e.g. number of compartments, sections,
synapses, etc.) but one can expect 4-7x reduction with the use of CoreNEURON. Note that \emph{CoreNEURON Online mode}
will need \( \frac{1}{7} \)x to \( \frac{1}{4} \)x more memory during the \emph{Memory Setup phase}. But once the model
is transferred to CoreNEURON for simulation, NEURON can free allocated memory.

Figure~\ref{fig:memory-reduction} on the right shows the speedup achieved on single node for different models with
CoreNEURON compared to NEURON. Note that the Cortex and Hippocampus models are very large in terms of memory capacity
requirement. For single node performance analysis we used a smaller subset of these two models.

The memory layout and code vectorization optimization described in Section~\ref{subsection:data-layout-vectorisation}
shows greatest improvement when most of the computation time is spent in channel and synapse computations. The Cortex,
Cortex+Plasticity and Hippocampus models have cells with 200 to 800 compartments and ~20 different channel types. This
makes these models compute intensive and get benefited most with CoreNEURON. The Ring network model has computations
only from passive dendrites and active soma.

Intel KNL has 512-bit SIMD vectors and high bandwidth memory (MCDRAM). One needs to efficiently utilize these hardware
features to achieve best performance. In the case of CoreNEURON, NMODL generated code is auto-vectorized by the compiler
and has \emph{SoA} memory layout to provide uniform, contiguous memory access. NEURON uses \emph{AoS} memory layout
which results in strided memory accesses. Due to the lower clock frequency of KNL cores, the performance impact of
non-vectorized code and strided memory accesses is high compared to other architectures. Hence CoreNEURON delivers
better performance on KNL compared to NEURON. Note that the Cortex+Plasticity and Hippocampus models have relatively
less improvement (2-4x) compared to the Cortex model (3-7x). This is because some of the channel and synapse
descriptions explicitly request integration methods that present compilers cannot efficiently vectorize. Alternative
code generation for these methods is being considered.

On the BlueGene/Q platform the speedup with most of the models is limited to 2x. This is because the IBM XL compiler is
not able to vectorize most of the channel and synapse kernels. Observed performance improvement on this platform is due
to more efficient memory accesses from the \emph{SoA} layout discussed in the
Section~\ref{subsection:data-layout-vectorisation}.

GPU support has been recently added to CoreNEURON and not all models are adapted for GPU yet. The Ring network model has
large number of identical cells which suits SIMD computations on GPU and hence shows significant performance improvement
compared to other architectures. The Traub model has a small number of cells exposing limited parallelism and the
Dentate model has gap junctions which require copying of voltages between CPU and GPU every timestep. This limits the
performance improvement on GPU.

\begin{figure*}[t]
    \centering
    \includegraphics[width=1\textwidth]{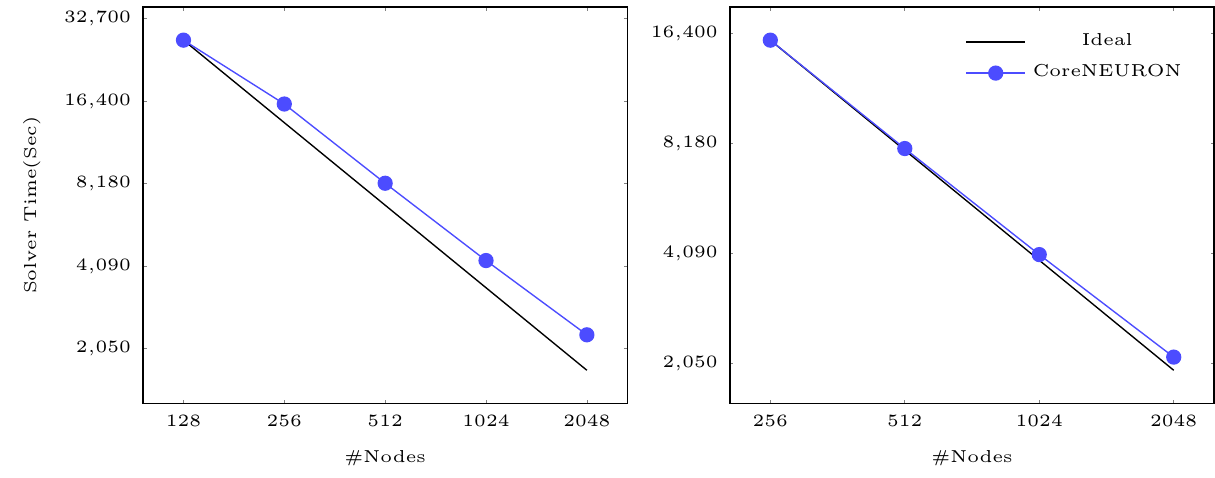}
    \caption{Strong scaling of CoreNEURON on BB4 system for Cortex+Plasticity model on the left and Hippocampus CA1 model model on the right}
    \label{plot:hippocampus-plasticity-strong-scaling}
\end{figure*}

Due to the large memory requirement of Cortex + Plasticity and Hippocampus models, a minimum of 2048 nodes of the BB4
system are required when NEURON is used. By using \emph{CoreNEURON Offline Mode}, users can now simulate the
Cortex+Plasticity model with 128 nodes and the Hippocampus model with 256 nodes of BB4. This is a significant usability
improvement because of limited compute partition size and long job queues on this shared computing resource.
Figure~\ref{plot:hippocampus-plasticity-strong-scaling} shows strong scaling of Cortex+Plasticity and Hippocampus models
simulating one second of biological time on the BB4 system with CoreNEURON. As these models are compute intensive and a
small fraction of execution time is spent in spike communication, the scaling behavior depends on how well a given
number of cells can be distributed across the available number of ranks to yield good load balance. Both models show
excellent strong scaling behavior up to 2,048 nodes. Due to the large size range of morpho-electrical neuron types, at
least 7-10 cells per MPI process are required to achieve good load balance. With 32,000 MPI processes (16 ranks per node)
and about 219,000 cells of Cortex+Plasticity, the load balance is not as good as with the Hippocampus model of about 789,000 cells.
Hence, the Cortex+Plasticity model exhibits poorer scaling behavior compared to the Hippocampus model.

\section{Discussion}
\label{section:conclusion-future-work}
Modern compute architectures can significantly boost application performance and the study of the brain in silico is in
dire need to embrace this capability and trend. Accordingly, the widely used NEURON simulator that supports a large
variety of models has been over the years successfully adapted to embrace massively parallel architectures, but its
primary design goals were to allow for a flexible definition of models and interactive introspection thereof. It was
neither designed for ultimate memory efficiency nor maximal performance. However, the larger and more detailed the
models get, the larger are the resource requirements to simulate those models. Eventually, the costs of a system required for an
un-optimized simulator should be weighed against the effort of reworking the simulator to make more efficient use of resources. In the
context of the Blue Brain Project we took the decision to contribute to making the NEURON simulator more efficient for
large models, effectively leading to reduced resource requirements, faster time-to-solution, or simply the capability to
run bigger models on a given resource.

{\bf Compatibility with existing NEURON models} \newline As the neuroscience community has developed and shared thousands of
models with NEURON, compatibility and reproducibility has been one of the primary design goals. To maintain maximal
compatibility, we chose the path of extracting the computational relevant parts of NEURON into a library called
CoreNEURON and adapting it to exploit the computational features of modern compute architectures. This is a different
path as for example taken by the Arbor initiative~\cite{arbor:github},~\cite{klijn:836542} which started its
developments from scratch. While such a fresh start has its benefits in terms of designing for future architectures from
the start, we can show that the transformation approach we took immediately gives compatibility with a large number of
existing NEURON models with minimal modification. Currently, CoreNEURON does not handle non thread-safe models and requires
NMODL modifications if constructs like POINTER~\cite{nmodl:pointer} are used. We are working on handling such models
transparently.

{\bf Flexibility for Model Building and Efficiency for Model Simulation} \newline Many modeling workflows related to detailed
brain models require flexibility for quickly inspecting and changing the models. By extracting the compute engine from
the NEURON simulator environment and providing different methods of how it can interact with the NEURON simulator, one
maintains the flexibility of NEURON for the construction of the models and can more easily apply optimizations to the
compute engine for the costly simulation phase. The \emph{Offline} execution mode of CoreNEURON provides flexibility
to build and simulate large network models that cannot be simulated with NEURON. Thanks to the use of MPI, and the OpenMP
and OpenACC programming models to achieve portability across different architectures such as multi-core, many-core
CPUs and GPUs.

{\bf Reduced Memory and Faster Time-to-Solution} \newline The data structure changes allow CoreNEURON to use significantly
less memory compared to NEURON. The \emph{SoA} memory layout and code vectorisation allow CoreNEURON to simulate models
efficiently. We benchmarked five different network models on different architectures showing 4-7x memory usage
reduction and 2-7x execution time improvement.

{\bf Future Work} \newline We discussed the implementation of the most significant changes and optimizations in NEURON and CoreNEURON.
Although CoreNEURON can be used transparently within NEURON, users cannot currently access or modify
model properties during integration. Work is ongoing in regard to bidirectional data copy routines activated by normal
NEURON variable name evaluation and assignment syntax ranging in granularity from the entire model, to specific named
arrays, down to individual variables. On the numerical side, CoreNEURON today supports network simulations using the
fixed time step method but not the variable time step integration method (CVODE)~\cite{cvode:cohen1996}. The latter is
rarely used in network simulations because state or parameter discontinuities in response to synaptic events demand
continuous re-initialization of variable step integrators. Research is ongoing on how to improve the applicability of
variable time step schemes in network simulation and can be considered for inclusion at a later stage. Currently,
mapping of multiple MPI ranks to GPUs is not optimal and this will be addressed in future releases. Lastly, the NMODL
source-to-source translator will be improved to generate efficient code for stiff, coupled, nonlinear gating state
complexes that require the \emph{derivimplicit} integration method as well as the generation of optimal code for GPUs.

{\bf Availability} \newline CoreNEURON and code generation program MOD2C are open sourced and available on
GitHub\cite{mod2c:github},~\cite{CoreNeuronGitHub}.

\section*{Acknowledgements}

This work has been funded by the EPFL Blue Brain Project (funded by the Swiss ETH board), NIH grant number R01NS11613 (Yale University), the European Union Seventh Framework Program (FP7/2007Â­2013) under grant agreement n$^\circ$ 604102 (HBP) and the European Union's Horizon 2020 Framework Programme for Research and Innovation under Grant Agreement n$^\circ$ 720270 (Human Brain Project SGA1) and Grant Agreement n$^\circ$ 785907 (Human Brain Project SGA2). The authors would like to thank Bruno Magalhaes, Francesco Cremonesi, Sam Yates and Timothee Ewart for contributions to the CoreNEURON development.

\printbibliography

\end{document}